%% file: main.tex
\documentclass[a4paper]{spie}  

\newcommand\degr{\hbox{$^\circ$}}

\usepackage{amsmath,amsfonts,amssymb}

\usepackage{graphicx}
\usepackage[colorlinks=true, allcolors=blue]{hyperref}
\usepackage{textgreek} 
\usepackage{support-caption}
\usepackage{subcaption}
\usepackage{rotating}
\usepackage{aasmacros}

\title{The MICADO Atmospheric Dispersion Corrector: Optomechanical design, expected performance and calibration techniques}
\author[a,b,c]{J.A.~van~den~Born}
\author[a]{R.~Romp}
\author[a]{A.W.~Janssen}
\author[a]{R.~Navarro}
\author[b,d]{W.~Jellema}
\author[b]{E.~Tolstoy}
\author[c]{B.~Jayawardhana}
\author[e]{M.~Hartl}

\affil[a]{\footnotesize NOVA Optical Infrared Instrumentation Group at ASTRON, Oude Hoogeveensedijk 4, NL-7991 PD Dwingeloo, the Netherlands}
\affil[b]{\footnotesize Kapteyn Astronomical Institute, University of Groningen, PO Box 800, NL-9700 AV Groningen, the Netherlands}
\affil[c]{\footnotesize Engineering and Technology Institute Groningen, University of Groningen, Nijenborgh 4, NL-9747 AG Groningen, the Netherlands}
\affil[d]{\footnotesize SRON Netherlands Insitute for Space Research, PO Box 800, NL-9700 AV Groningen, the Netherlands}
\affil[e]{\footnotesize Max-Planck-Institut f\"ur extraterrestrische Physik (MPE), Giessenbachstr. 1, D-85748 Garching, Germany}

\authorinfo{E-mail: born@astro.rug.nl}

\begin{document} 
\maketitle

\begin{abstract}
The differential refraction of light passing through the atmosphere can have a severe impact on image quality if no atmospheric dispersion corrector (ADC) is used. For the Extremely Large Telescope (ELT) this holds true well into the infrared. MICADO, the near-infrared imaging camera for the ELT, will employ a cryogenic ADC consisting of two counter-rotating Amici prisms with diameters of 125 mm. The mechanism will reduce the atmospheric dispersion to below 2.5~milli~arcseconds (mas), with a set goal of 1~mas. In this report, we provide an overview of the current status of the ADC in development for MICADO. We summarise the optomechanical design and discuss how the cryogenic environment impacts the performance. We will also discuss our plan to use a diffraction mask in the cold pupil to calibrate and validate the performance once the instrument is fully integrated.
\end{abstract}



\input{01_introduction}
\input{02_optical_design}

\input{03_mechanical_design}

\input{05_performance_validation}
\input{06_conclusion}


\newpage
\acknowledgments 
We would like to acknowledge the Leids Kerkhoven-Bosscha Fund (LKBF) for partially funding our presence at this conference. We would also like to thank the SPIE Student Conference Support for allowing the first author to present his work here.

Furthermore, we would like to thank Niels Tromp and Gert Musters for their contributions to the early conceptual design and their work on the MICADO ADC Mechanism prototype. Finally, we would like to acknowledge the rest of the MICADO team at NOVA for their help in design, assembly, integration, testing, programming and manufacturing of the prototype and final hardware.

\bibliography{report} 
\bibliographystyle{spiebib} 

\end{document}

%% file: 01_introduction.tex
\section{INTRODUCTION}
\label{sec:introduction}
\begin{figure}
\begin{minipage}[t]{.47\textwidth}
\centering
\captionsetup{width=\linewidth}
\includegraphics[width=\linewidth]{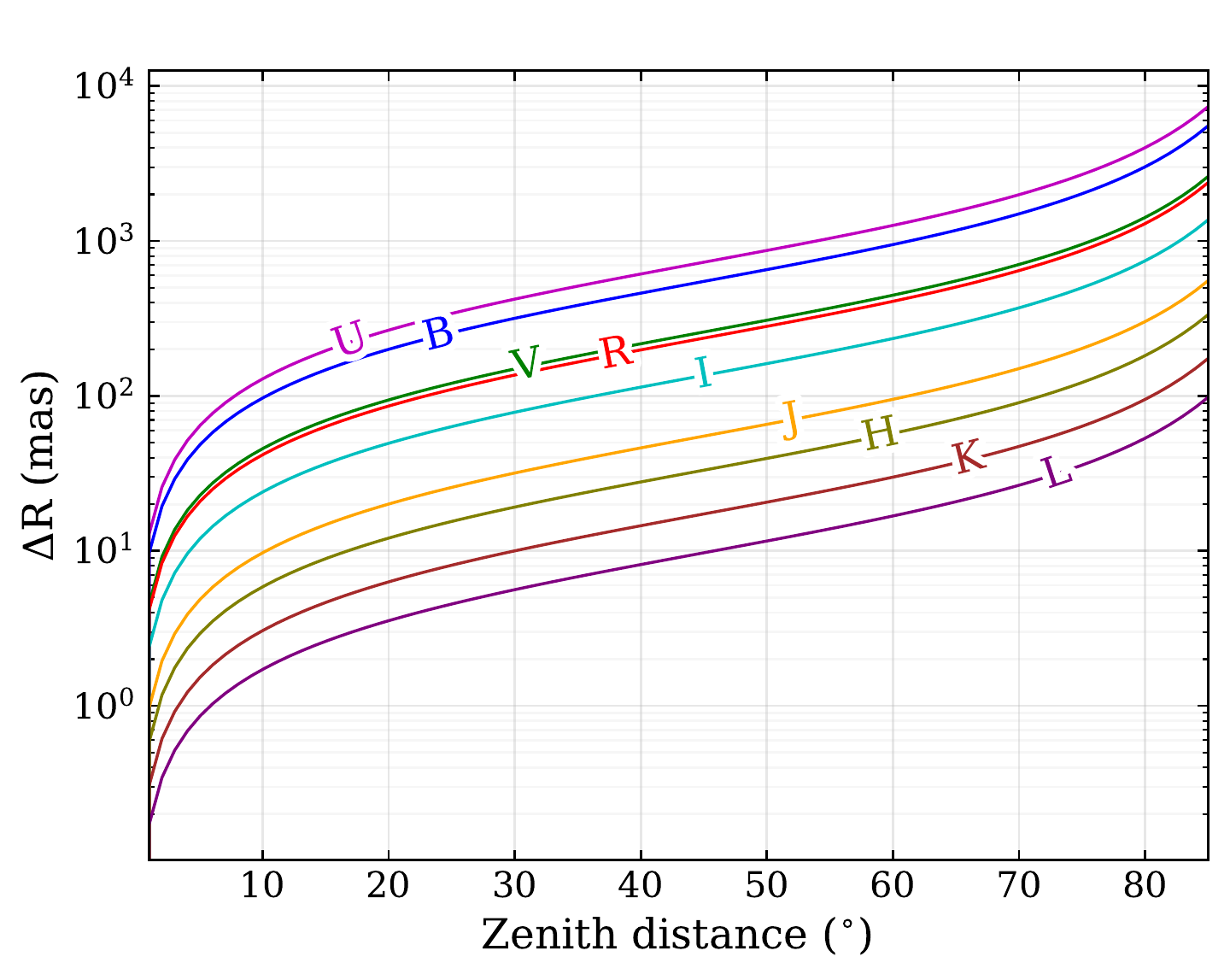}
\captionof{figure}{The magnitude of PSF elongation caused by atmospheric dispersion increases for shorter wavelengths as well as with increasing zenith distance. Shown here are typical values for the atmospheric dispersion in milli arcseconds. Assumed here are atmospheric conditions representative of Cerro Armazones, the future site of the ELT.}
\label{fig:atmospheric dispersion}
\end{minipage}%
\hfill
\begin{minipage}[t]{.47\textwidth}
\centering
\captionsetup{width=\linewidth}
\includegraphics[width=\linewidth]{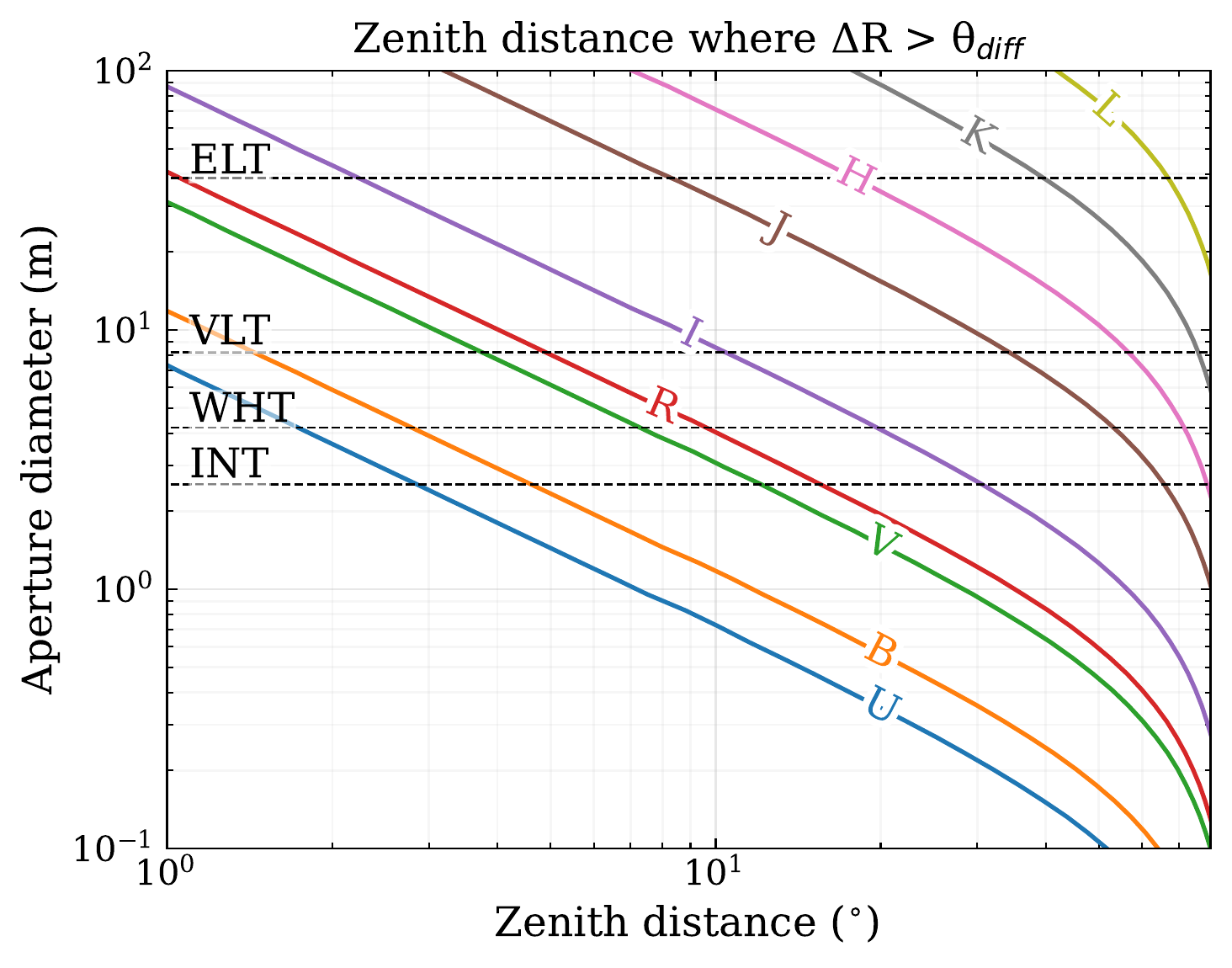}
\captionof{figure}{The increased resolution allowed for by an ELT class telescope makes an atmospheric dispersion corrector a definite requirement well into the near-infrared if diffraction limited performance is desired. In this figure, we can compare the atmospheric dispersion expected in several common passbands for a range of telescope aperture sizes. Again, atmospheric conditions similar to Cerro Paranal are assumed.}
\label{fig:atmospheric_dispersion_comparison}
\end{minipage}%
\end{figure}

Atmospheric dispersion is the differential refraction between light rays of different colour that are traversing the Earth's atmosphere. Its effect on a point source imaged by a telescope can be seen as an elongation of the point spread function (PSF) along the parallactic angle. The magnitude of the elongation scales with the wavelength and the observed zenith distance. This is apparent from the first order approximation describing the atmospheric dispersion,

\begin{equation}
    \Delta R = 206265 \times \left[ n(\lambda_1) - n(\lambda_2)\right] \tan{z}.\label{eq:plane-parallel}
\end{equation}
Here $\Delta R$ denotes the atmospheric dispersion and is given in arcseconds, $n(\lambda)$ is the refractive index of air at some wavelength $\lambda$ and $z$ is the angular distance from zenith. A comparison of the atmospheric dispersion at different passbands is shown in Fig.~\ref{fig:atmospheric dispersion}.

For small or seeing limited telescopes, atmospheric dispersion is generally not a major problem. However, with the increase in resolution offered by the upcoming generation of extremely large telescopes and with adaptive optics becoming commonplace, this effect must be corrected for. The severity of this problem is illustrated in Fig.~\ref{fig:atmospheric_dispersion_comparison}, where we use eq.~\eqref{eq:plane-parallel} to show the zenith distance at which the atmospheric dispersion starts to exceed the diffraction limit of a telescope with a given primary mirror diameter. Assuming that the adaptive optics systems on a 30-meter class telescope work well, the atmospheric dispersion adversely affects image quality even in \textit{K}-band, at a moderate 45\degr\ from zenith. At shorter wavelengths, such as \textit{J}-band, almost all observations will be adversely affected, as the atmospheric dispersion in \textit{J}-band at 10\degr\ from zenith already exceeds the diffraction limit.

Thus, it should be no surprise that MICADO, the Multi-AO Imaging CamerA for Deep Observations \cite{Davies2022}, comes equipped with an atmospheric dispersion corrector (ADC). MICADO is the near-infrared imager in development for the Extremely Large Telescope. The state-of-the-art adaptive optics systems and the ADC will enable the instrument to deliver diffraction limited performance from 0.8 to 2.4 \textmu m\cite{MAORY2016, MAORY-SCAO2019}.

To obtain the desired image quality, photometric performance and astrometric accuracy, the instrument design requires that the chromatic dispersion is corrected to a level below 2.5 milli arcseconds (mas) in \textit{J}-, \textit{H}- and \textit{K\textsubscript{s}}-band. The goal is to reach a correction accuracy of 1.0 mas. Although it is not a requirement for instrument acceptance, a level of residual dispersion below 0.4 mas is desired for optimum astrometric performance of the instrument. However, it seems likely that this last value will only be obtainable for a limited subset of the possible observing conditions.

Besides the high level of dispersion correction, the MICADO ADC will also be one of the first ADCs built for a cryogenic environment, the other being the IRIS ADC in development for the Thirty Meter Telescope\cite{Phillips2016}. In such a challenging environment, changes in material properties and increases in the wear and tear should be carefully considered during the design phase. Not in the least because opening the instrument cryostat will be costly in terms of lost observing time. We will discuss some the challenges and our solutions in the following sections.

This report will provide an overview of the current status of the MICADO atmospheric dispersion corrector, in development at NOVA. In sections~\ref{sec:opticaldesign} and \ref{sec:mechanicaldesign} we will discuss the optical and mechanical design of the ADC, respectively. Then, in section~\ref{sec:calibration} we will discuss how we plan to validate the performance as the instrument comes online at the end of 2027. We conclude this report with some summarising thoughts in section~\ref{sec:conclusion}.

%% file: 02_optical_design.tex
\section{OPTICAL DESIGN}\label{sec:opticaldesign}
\begin{figure}
    \centering
    \includegraphics[width=0.55\linewidth]{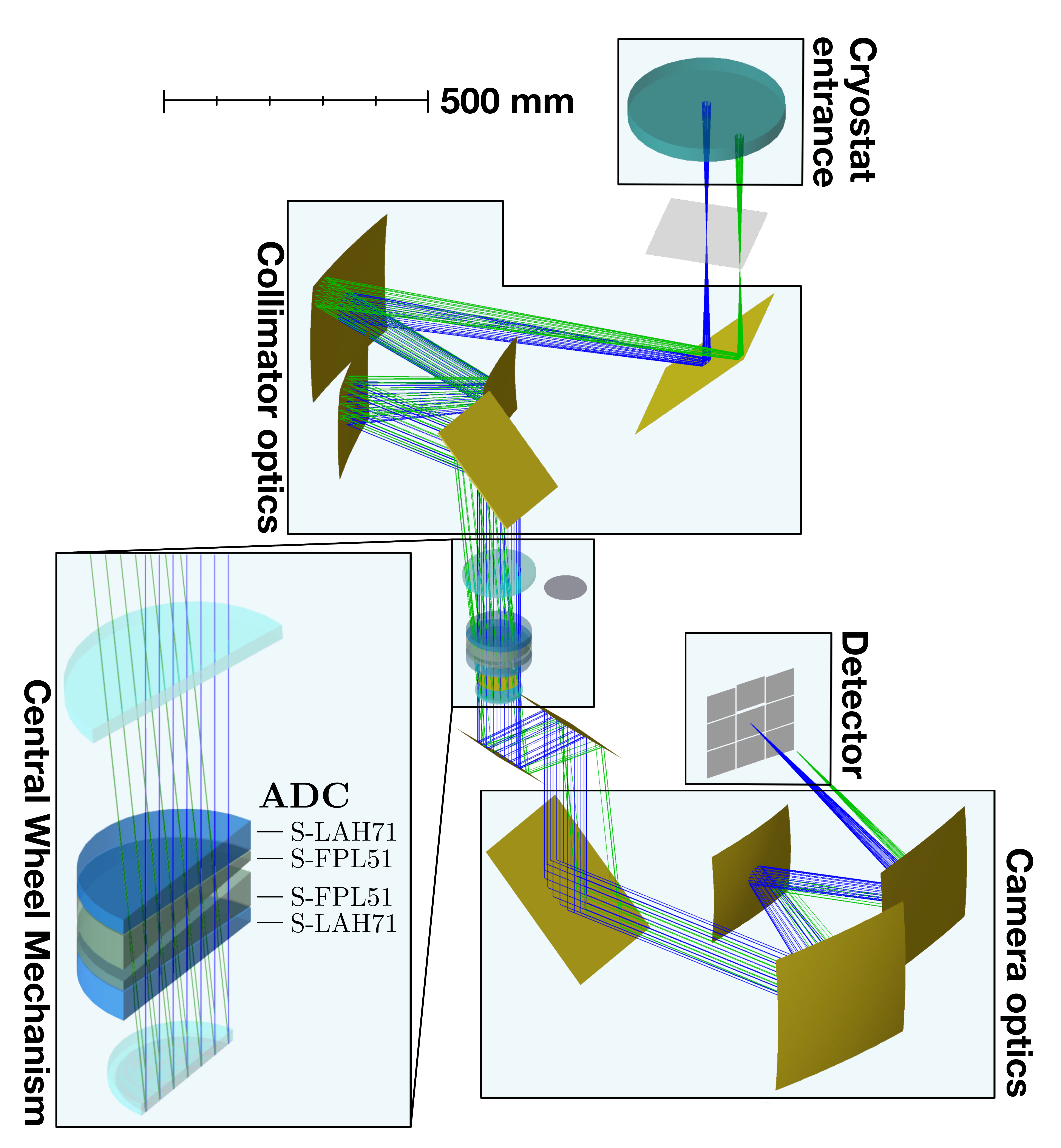}
    \caption{Overview of the optical design of the MICADO cold optics, shown here in the standard imaging mode. The ADC is part of the central wheel mechanism and is located near the optical pupil, between the passband filters and pupil optics.}
    \label{fig:MICADO_Optical_Design}
\end{figure}

The MICADO optical design comprises of three main components, as illustrated in Fig.~\ref{fig:MICADO_Optical_Design}. First, the collimator optics shape the light from the telescope into a collimated light beam. Then, the filter wheels, ADC and pupil optics are used to modify the light characteristics according to the observation goals. Finally, the collimated light gets imaged onto the focal plane through the camera optics.

The MICADO ADC consists of two sets of counter-rotating Amici prisms, located near the cold pupil of the instrument. Each set contains one S-FPL51 and one S-LAH71 prism. The prisms are 125 mm in diameter and have a thickness of 15 mm at the centre. 

The refractive indices of the glass materials and the apex angles of the prisms combine to emulate the dispersive properties of the Earth's atmosphere. By rotating the prisms along the optical axis, the effective apex angle along the direction of dispersion is varied and tuned such that the dispersion by the ADC approximates the atmospheric dispersion, but is in opposite direction. Because the effective refractivity of the ADC does not exactly equal that of the atmosphere, the atmospheric dispersion can only be perfectly corrected at two wavelengths at a time. Triplet designs improve upon this by adding more degrees of freedom in the design \cite{Kopon2008}, but some amount of residual dispersion will always remain within a selected passband.

When the optical design is fixed, then the optimum angular position of the prisms can be approximated from the prism geometry, under the assumptions of a plane-parallel atmosphere, an ADC located in the pupil, and small angles of incidence\cite{vandenBorn2020a, Egner2010}.
\begin{equation}
    \theta_{\textnormal{opt}} = \cos^{-1}\big(c_f \tan z\big), \label{eq:optPos}
\end{equation}
Here, $z$ denotes the observed distance from zenith and $c_f$ is a constant that changes for every passband. This expression requires that a non-dispersive configuration is obtained when $\theta = 0$. For any other value, both prisms must rotate in opposite directions by the computed value $\theta_{\textnormal{opt}}$. In Ref.~\citenum{vandenBorn2020a}, the filter constant for a two doublet ADC design was derived to be 
\begin{equation}
    c_f = \frac{D_{\textnormal{\tiny{EPD}}}}{D_{\textnormal{\tiny{ADC}}}} \frac{\Delta n_{\textnormal{atm}}}{\Delta n_g (A_1 + A_2)}, \label{eq:filter_constant}
\end{equation}
where $D_{\textnormal{\tiny{EPD}}}$ and $D_{\textnormal{\tiny{ADC}}}$ are the diameter of the beam footprint at the entrance pupil and the ADC respectively. Then, $\Delta n_{\textnormal{atm}}$ and $\Delta n_g$ refer to the difference in refractive index of the two wavelengths of interest for the atmosphere and the glass. Finally, $A_1$ and $A_2$ are the apex angles of the two prisms making up the Amici prism.

The glass optimisation for the MICADO ADC was done through a deterministic first order dispersion analysis of all glass materials in the OHARA, SCHOTT and Infrared catalogues. The search did exclude some specific unwanted materials and also included moderate constraints on the allowed surface angles. An optimum combination was found in S-FPL51 and S-LAH, in agreement with the glass optimisation done for the IRIS ADC\cite{Phillips2016}. Following this, an optimisation of the prism apex angles was done in ZEMAX OpticStudio, within the complete optical design of the MICADO cold optics, leading to the current nominal optical prescription of the ADC, as provided in Table~\ref{tab:adc_optical_prescription}. Later, during the procurement of the chosen glass, a better material combination was found in S-LAH71 and S-FPM4. The improvement in performance was only minimal and the effort to incorporate it into the design was considered not worth the potential gains in this stage of the project. But it may be worth consideration in future instruments.

\begin{table}
    \captionsetup{width=0.85\linewidth}
    \caption{Optical prescription of the ADC. The individual Amici prisms are of an equal, but mirrored design. The individual glass prisms within a pair will include a gap of approximately 1 mm, which is not shown in this table.}
    \label{tab:adc_optical_prescription}
    \centering
    \begin{tabular}{cccc}
    \hline
    Surface & Thickness (mm) & Glass & Tilt angle (\degr) \\
    \hline
    1 & 15  & S-LAH71 & 1.6\\
    2 & 15  & S-FPL51 & 6.3\\
    3 &  8 & - & -1.5\\
    4 & 15 & S-FPL51 & -1.5\\
    5 & 15 & S-LAH71 & 6.3\\
    6 & - & - & 1.6\\
    \hline
\end{tabular}
\end{table}

\begin{figure}[htpb]
    \centering
    \begin{subfigure}[b]{0.49\textwidth}
        \centering
        \includegraphics[width=\textwidth]{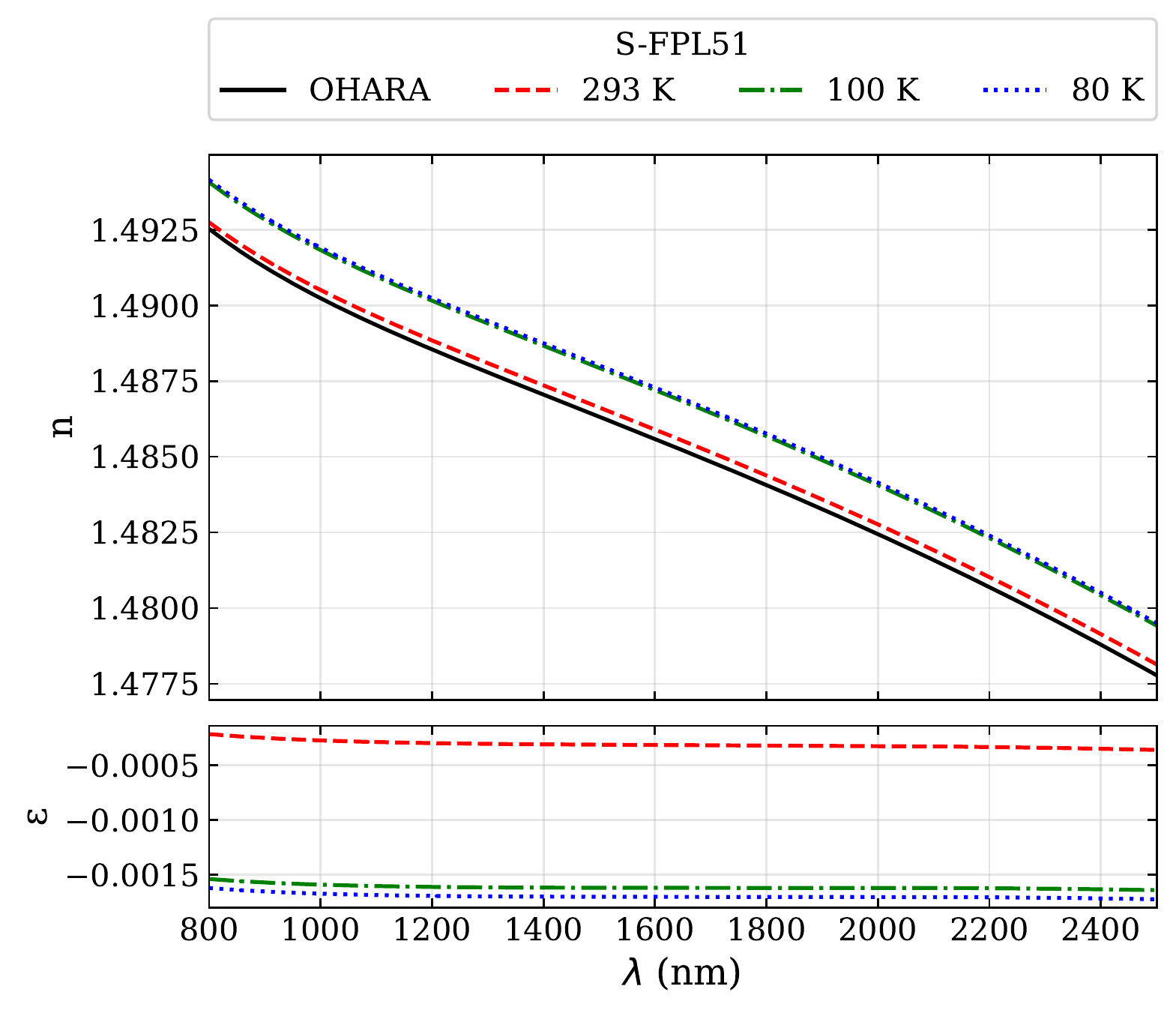}
    \end{subfigure}
    \hfill
    \begin{subfigure}[b]{0.49\textwidth}  
        \centering 
        \includegraphics[width=\textwidth]{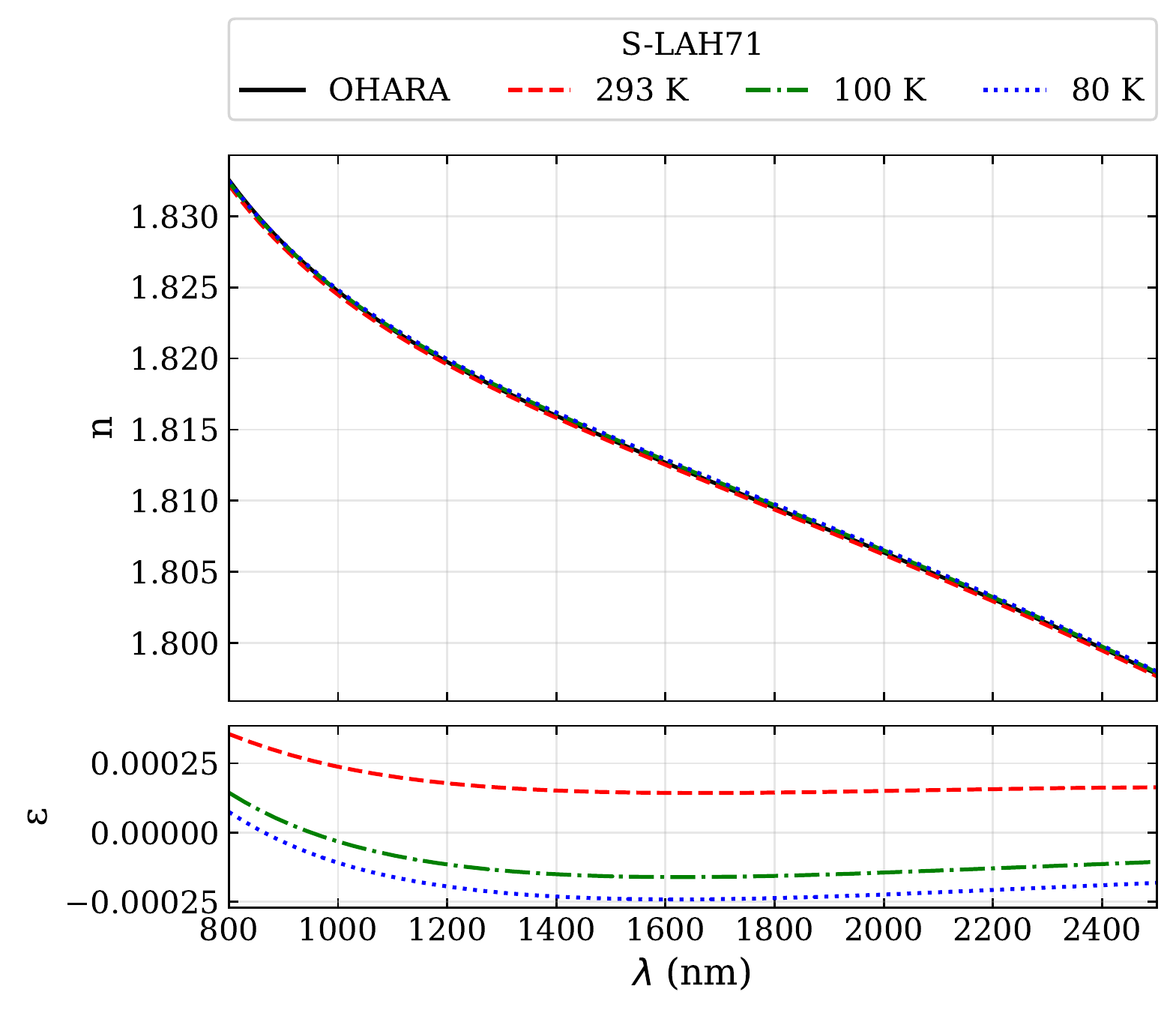}
    \end{subfigure}
    \caption{The index of refraction of the S-FPL51 (left) and S-LAH71 (right) materials that will be used in the final hardware at 293~K, 100~K and 80~K. The black line denotes the curve given by OHARA in the data sheet. The bottom panels show the deviation $\varepsilon$ of our samples from the data sheets.}\label{fig:measured_refractive_indices}
\end{figure}

\begin{figure}
    \centering
    \includegraphics[width=0.6\linewidth]{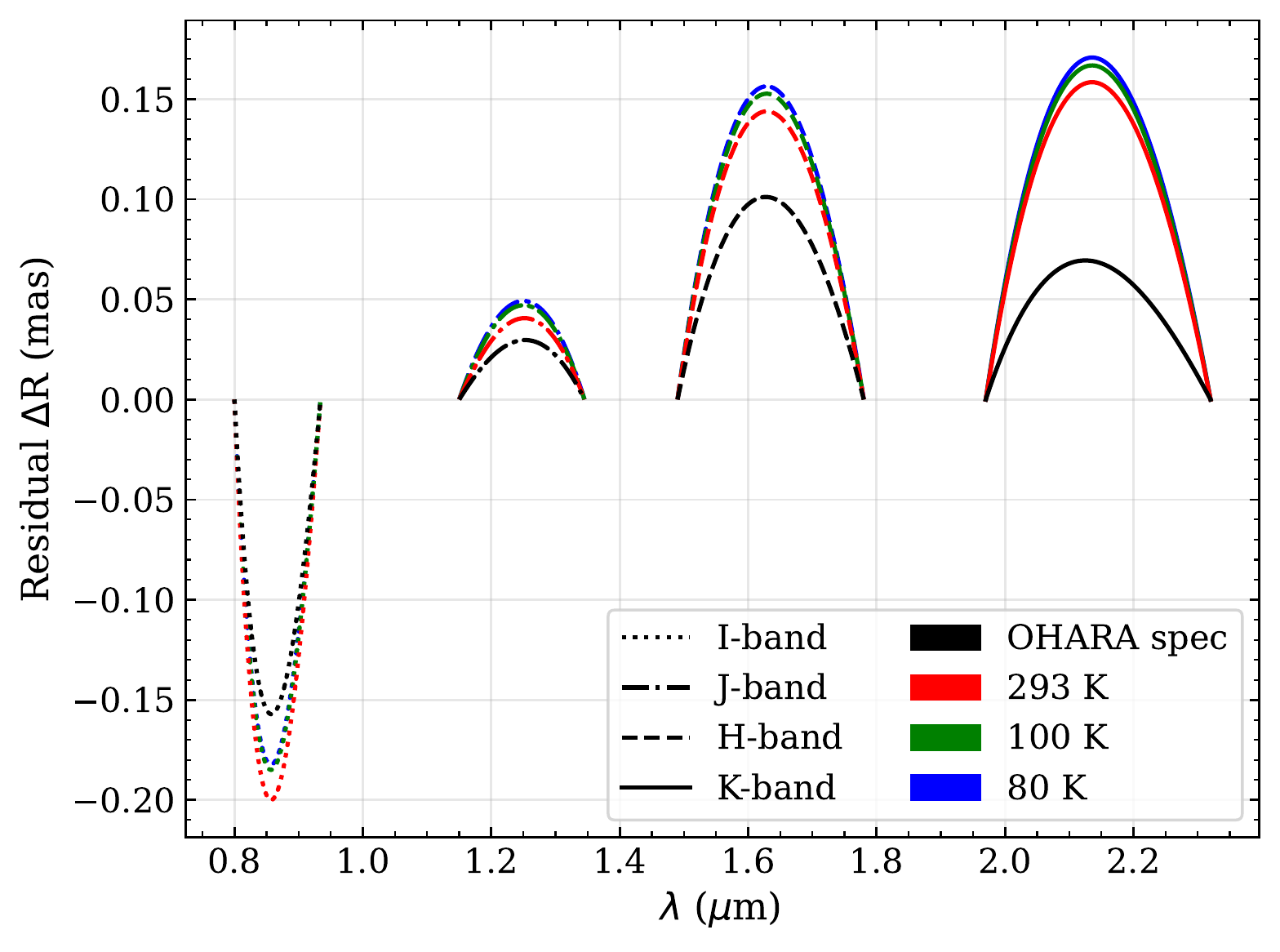}
    \caption{The ADC can only be optimised to perfectly correct the dispersion at two wavelengths simultaneously. If we assume that the positioning of the prisms is such that the edges of the desired passband are perfectly corrected, then we can inspect the intraband residual dispersion. In this figure, we have taken the nominal ADC design and optimised the ADC for the edge wavelengths of the four major passbands, here for an observation at 30\degr\ from zenith. The colored lines then give the intraband dispersion for the measured refractive indices at 293, 100 and 80 K, respectively. The effect will scale with the observed zenith angle. For example, the presented magnitude will triple when observing at 60\degr from zenith. This figure is adapted from Ref.~\citenum{vandenBorn2020a}.}
    \label{fig:measured_n_ADC_performance}
\end{figure}

The optimum apex angles are heavily dependent on the refractive index of the sourced material. Slight variations between the refractive index of a material in different manufacturing runs can lead to a deviation from the expected performance. Therefore, the refractive indices of the produced S-FPL51 and S-LAH71 substrates have recently been measured by OMT Solutions at room temperature and in a representative cryogenic environment. The results are shown in Fig.~\ref{fig:measured_refractive_indices}. For S-FPL51, we find that the refractivity of our sample is offset by $1.5\times10^{-3}$ at to room temperature compared to the values provided by OHARA. For S-LAH71 the difference is less at approximately $2.5\times10^{-4}$. Figure~\ref{fig:measured_n_ADC_performance} illustrates the loss in performance with respect to the residual dispersion within a given passband for the measured refractive indices. With a new optimisation of the apex angles, the nominal design performance can be regained. This will be done in the near future.

The intraband residual dispersion is the one aspect of performance that is most affected by the change in refractive index of the glass. Another performance metric is the maximum zenith distance at which the ADC can correct for the atmospheric dispersion. The measured refractive indices change this metric by a small amount, although its exact value differs for the different major bands in the instrument. For example, in \textit{I}-band the measured refractive indices decrease the maximum dispersion that the ADC can correct for by 5~mas, which decreases the maximum zenith distance at which the ADC can obtain an optimum correction by half a degree to 64.1\degr. In \textit{H}-band the maximum dispersion correction stays practically identical to the nominal design, while in \textit{K\textsubscript{s}}-band the maximum zenith distance at which the atmospheric dispersion can be fully corrected actually increases by 1.5\degr\ to 69\degr. Because observations at these large zenith distances do not drive the design of the ADC, or MICADO in general, we assign little weight to these numbers.

%% file: 03_mechanical_design.tex
\section{MECHANICAL DESIGN}\label{sec:mechanicaldesign}



\begin{figure}
    \centering
    \includegraphics[width=0.8\linewidth]{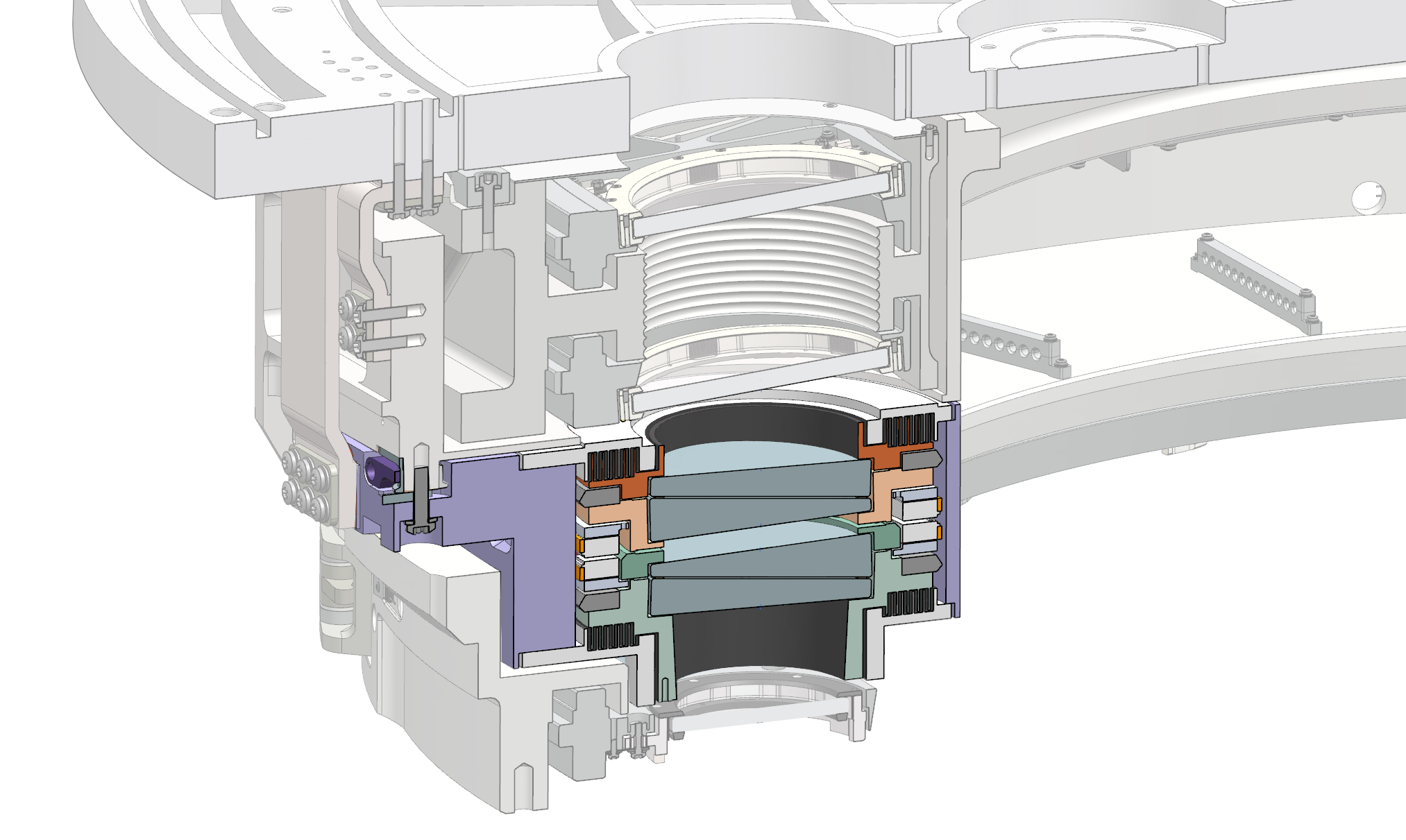}
    \caption{This cross-section view of the Central Wheel Mechanism (CWM) illustrates the location and general design of the ADC. The ADC is a separate module within the CWM, containing two similar mechanisms to rotate each of the two prism pairs. The height of the ADC module is 144 mm, and the full height of the design shown here, including the greyed out parts, is 400 mm . 
    }
    \label{fig:CWM_mech_design}
\end{figure}

The Central Wheel Mechanism subsystem of the MICADO instrument contains the filter wheels, pupil optics and the atmospheric dispersion corrector\cite{Romp2020}. Because all these components are ideally located at the optical pupil, the vertical design volume available for a suitable ADC mechanism is limited. An overview of the ADC within the CWM is provided in Fig.~\ref{fig:CWM_mech_design}. Additional constraints on the design work come from maximum acceptable thermal differentials between components, due to different values in the coefficient of thermal expansion, and strict requirements on wavefront quality, which limit the number of solutions to mount the optics. Furthermore, cryogenic mechanisms require careful consideration of wear and tear\cite{Krantz2014}, the behaviour of lubricants\cite{Roberts2012} and changing material properties.

To tackle the challenge of the limited physical volume, we have designed a friction drive mechanism to position the Amici prisms of the MICADO atmospheric dispersion corrector. This approach integrates the positioning and weight bearing of the ADC rotor into a single mechanism, while accommodating thermal expansion and minimising backlash.

It works as follows. The ADC consists of two nearly identical mechanisms, one for each prism set. In each mechanism the prisms are mounted to an Al-6082-T6 rotor wheel, coated with polytetrafluoroethylene (PTFE). This wheel is directly coupled to three stainless steel 15-5PH roller wheels, one of which is driven by a Phytron VSS43 stepper motor. The roller wheels are held in place in a cage with gold coated ball bearings. To account for changes in the physical size of the rotor wheel as it cools down or warms up, the powered roller wheel is spring loaded against the rotor wheel. The geometry of all wheels is designed such that the forces on the bottom and top of the roller wheels are equally distributed and the differential rotation velocity of the points of contact is negligible. The position of the prisms are determined by a 22-bit Zettlex IncOder absolute angular encoder. Because this encoder does not work at 77 K, it must be kept at an elevated temperature of 210 K. It is thermally isolated from the rest of the system to reduce thermal leakage and subsequently reduce the required heating power.

A friction drive is not the most obvious design choice with respect to the longevity of a cryogenic mechanism. In addition, NOVA did not have any experience with a cryogenic friction drive at the time this concept was first conceived. Therefore, a prototype was built to verify the concept, characterise the performance and update the design according to our findings. With this prototype, we have verified that the mechanism can be positioned with the desired accuracy and velocity, as well as that the tip, tilt, and the lateral run-out of the rotor wheel are within the instrument requirements. As we discovered, obtaining the ten year lifetime - preferably without maintenance - was the most difficult challenge of the test campaign. We designed an automated accelerated lifetime test, where the expected movements of the mechanisms during its ten years at the ELT were simulated in quick succession. The first attempt at the execution of this test resulted in mechanism that got stuck when significant wear increased the required torque to rotate the rotor to a level that could not be delivered by the stepper motor. Several test iterations were needed before a combination of roller and rotor wheel materials was found that could reach the equivalent of ten years of operation. Failed material combinations for the rotor and rollers included SS440B coated with MoST and Ti-6Al-4V, SS440B coated with MoST and SS15-5PH, Ti-6Al-4V and SS15-5PH. Finally, an Al6082-T6 rotor wheel coated with PTFE and SS15-5PH roller wheels showed very good performance during the full duration of the test, with no signs of imminent failure. There was, however, significant contamination of particulates. We are currently assessing how minimise this and to what extent we can control where the contamination collects within the mechanism.

\begin{figure}
    \centering
    \includegraphics[width=.75\linewidth]{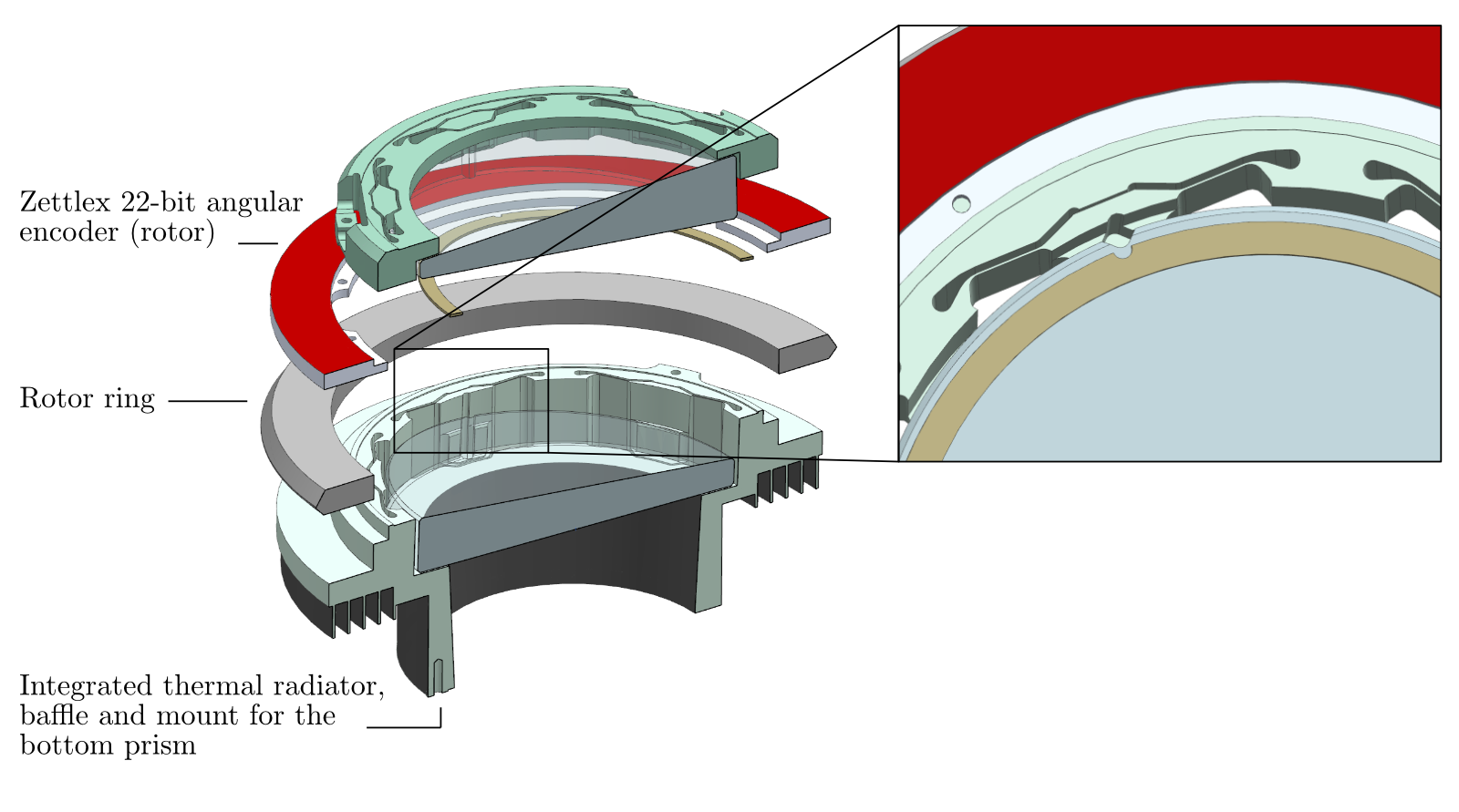}
    \caption{Six leaf springs hold each ADC prism in place, providing sufficient contact area for thermal transfer. To rotationally align the prisms, there is a notch in both the prism glass and the mount. The mount for the bottom prism forms the main structure of the rotor, with an integrated radiator. A flat ring (brown) creates a small gap with good parallelism between the two prisms. The top prism is then held in place by the mount cover (green), which also has six radially distributed leaf springs.}
    \label{fig:rotor_prism_mount}
\end{figure}

The prism mounting solution plays an important role in the design of the ADC. It shapes part of the alignment tolerances and it must be able to accommodate the thermal expansion differences between the glass and the rotor structure. The ADC prism mounts are using mostly the same design as the filter and pupil cells in the CWM\cite{Romp2020}, where six leaf springs provide sufficient clamping force to keep the glass in place, without introducing too large stresses to create significant birefringence effects. Because a glue joint between the two prisms within a doublet was deemed too risky with respect to stresses on the substrate surface during thermal cycles, the individual glass prisms will be held in place mechanically. A flat ring provides a small gap between the two prisms. A notch in the mount and the glass offers a way to align each prism rotationally during assembly. The prism mounts of the bottom ADC rotor are shown in Fig.~\ref{fig:rotor_prism_mount}.

\begin{figure}
    \centering
    \includegraphics[width=0.7\linewidth]{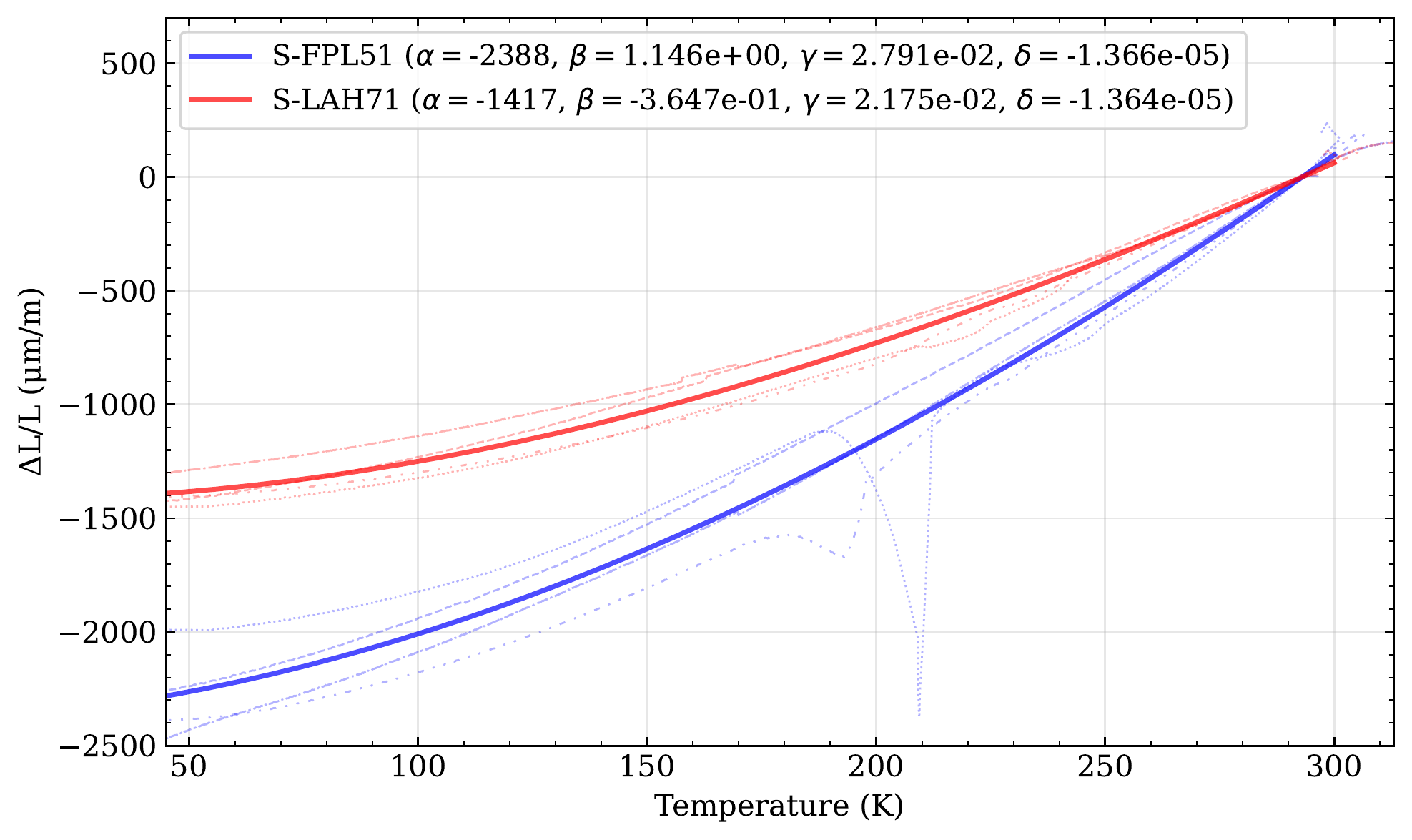}
    \caption{The measured linear coefficient of thermal expansion of the ADC glass materials. The thick lines represent the best fit curve of the two measurements, where each measurement is a full thermal cycle of the sample. The thin lines represent the individual measurements, consisting of two cooldowns and two warmups for each sample. The coefficients presented in the figure legend can be used to calculate the thermal expansion at any temperature relative to 293 K, following $\Delta L / L = \alpha + \beta T + \gamma T^2 + \delta T^3$. A temperature lag between temperature of the temperature sensor and the measured displacement was noted and explains the fairly large variation between the individual measurements.}
    \label{fig:measured_cte}
\end{figure}

Recently, we have measured the coefficient of thermal expansion of the S-FPL51 and S-LAH71 glasses, at temperatures between 50 K and room temperature. This measurement was done on samples of approximately 10 mm thick, from the same manufacturing batch as the final optical substrates. Using an Attocube IDS3010 interferometric displacement sensor, the change in the size of the samples was measured as a function of the temperature. Our results, shown in Fig.~\ref{fig:measured_cte}, indicate that the prisms will experience a radial shrinkage of 140 \textmu m and 80 \textmu m for S-FPL51 and S-LAH71, respectively. Along the optical axis, these numbers are 34 \textmu m and 19 \textmu m. Taking into account the expected thermal differentials and the expansion of the mount itself, the presented design will be a suitable solution.

At present, MICADO is at its Final Design Review\cite{Davies2022}. The atmospheric dispersion corrector has raised no major issues, although some work remains to fully detail out the design. With the information and experience we have obtained through the performed tests, we feel confident that the challenges described at the start of this section are solved or will be in the near future. 

%% file: 05_performance_validation.tex
\section{TECHNIQUE FOR PERFORMANCE VALIDATION}\label{sec:calibration}

Calibration and validation of the ADC performance will be an important part in the commissioning of MICADO. Current models describing atmospheric dispersion have not yet been tested at the level required for the instrument. Furthermore, extensive on-sky testing and optimisation of the ADC is time consuming and costly. Therefore, it is important to be able to verify the performance quickly and effectively. 

In Ref.~\citenum{vandenBorn2022}, a promising method was demonstrated on a small telescope, where a diffraction mask in the telescope pupil was used to magnify the PSF elongation caused by the atmospheric dispersion. The mask acts as a diffraction grating to create artificial broadband speckles around the core of the PSF. When chromatic elongation is present in the point spread function, either caused by the atmosphere or by internal optics, the shape and direction of the artificial speckles is affected. A common intersection point of the speckle pointing directions, called the radiation centre, exists and provides a measure for the dispersion present in the image. The dispersion $\Delta R$ can be retrieved by measuring the distance between the radiation centre and the PSF core $d_{\textnormal{RC}}$, using
\begin{equation}
    \Delta R = \left( \frac{\lambda_2 - \lambda_1}{\lambda_2}\right) d_{\textnormal{RC}}. \label{eq:atm_disp_measure}
\end{equation}
In this equation, $\lambda_1$ and $\lambda_2$ are the cut-on and cut-off wavelengths of the used passband.


In this section, we will discuss how this principle can be applied to the MICADO ADC and if this is a suitable method for the performance validation during and after commissioning of the instrument.

\subsection{Design of a calibration mask}
The two primary design parameters for the ADC calibration mask are the number of artificial speckles that are desired for the image analysis and the distance at which these speckles should appear from the core of the PSF. 
The first is determined by the n-fold rotational symmetry of the mask pattern or by the number of orientations in which a grating pattern appears in different pupil zones. For example, a classic Bahtinov mask, as used in Ref.~\citenum{vandenBorn2022}, creates six speckles, and a checkerboard pattern creates four speckles around the central star. The second design paramater, the angular distance between the speckles and the central star, is determined by the number of periods over the pupil diameter, following 
\begin{equation}
    \theta = N \frac{m\lambda}{D},
\end{equation}
where $N$ is the number of periods, $m$ is the diffraction order and $D$ is the entrance pupil diameter of the telescope.

Using these two principles, we have explored six mask designs. The first three are inspired by the Bahtinov masks often used in amateur astronomy, where the grating patterns consist of straight lines. The other three are small repeating apertures, that fill the pupil. In all cases, we assume binary amplitude modulation and we have tuned the individual masks such that they have similar overall transmission for a valid comparison. The number of periods over the pupil diameter was chosen at 28, so that the first diffraction order falls within the correction radius of the adaptive optics system in the major passbands of MICADO. If applicable, the mask design is rotated such that the diffraction pattern does not overlap with the diffraction spikes of the telescope structure. This increases the contrast of the speckles relative to the local image background.

\begin{figure}
    \centering
    \includegraphics[width=\linewidth]{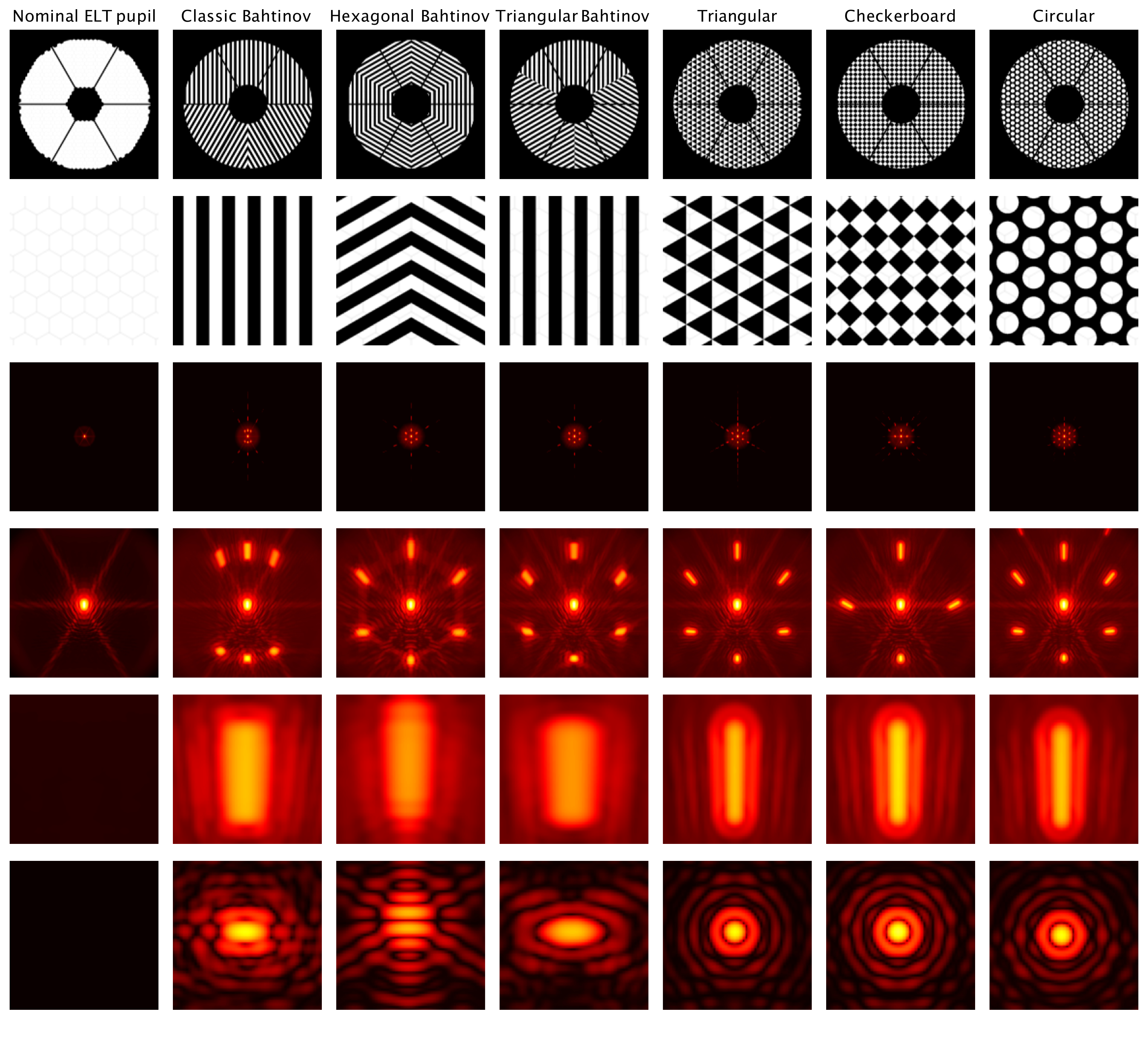}
    \caption{A comparison of the \textit{K\textsubscript{s}}-band PSF for the nominal ELT pupil and six mask designs. The simulation assumes an observation of a natural guide star and seeing of 0.5 arcsec. The rows in this figure show, from top to bottom, the full pupil, a closeup of the pupil pattern, an extended view of the PSF, a close up of the PSF with the first order diffraction speckles and then a close up of the first order diffraction speckle itself. The last row shows the monochromatic speckle, where $\lambda = 1.97$ \textmu m. The intensity in the panels is normalised along each row, but not from row to row.}
    \label{fig:mask_design_comparison_monochromatic}
\end{figure}

We have simulated the PSF of MICADO with these diffraction masks using the the Fourier optics framework of Ref.~\citenum{vandenBorn2020b}, including an extension to the framework of an analytical treatment of Single Conjugate Adaptive Optics (SCAO) based on \texttt{anisocado}\cite{anisocado2020}. Figure~\ref{fig:mask_design_comparison_monochromatic} illustrates that the small repeating apertures provide higher contrast in the diffraction speckles. The Bahtinov-like masks have a more elongated speckle shape, with the \textit{Hexagonal Bahtinov} mask design even experiencing severe self-interference, similar to the many slit experiment. The triangular apertures design is the preferred design here due to the aforementioned higher contrast in the speckles and because all higher order diffraction occurs along well defined axes, decreasing the likelihood of source confusion.

\subsection{Simulation of the performance validation accuracy}\label{subsec:validation}
\begin{figure}
    \centering
    \includegraphics[width=0.65\linewidth]{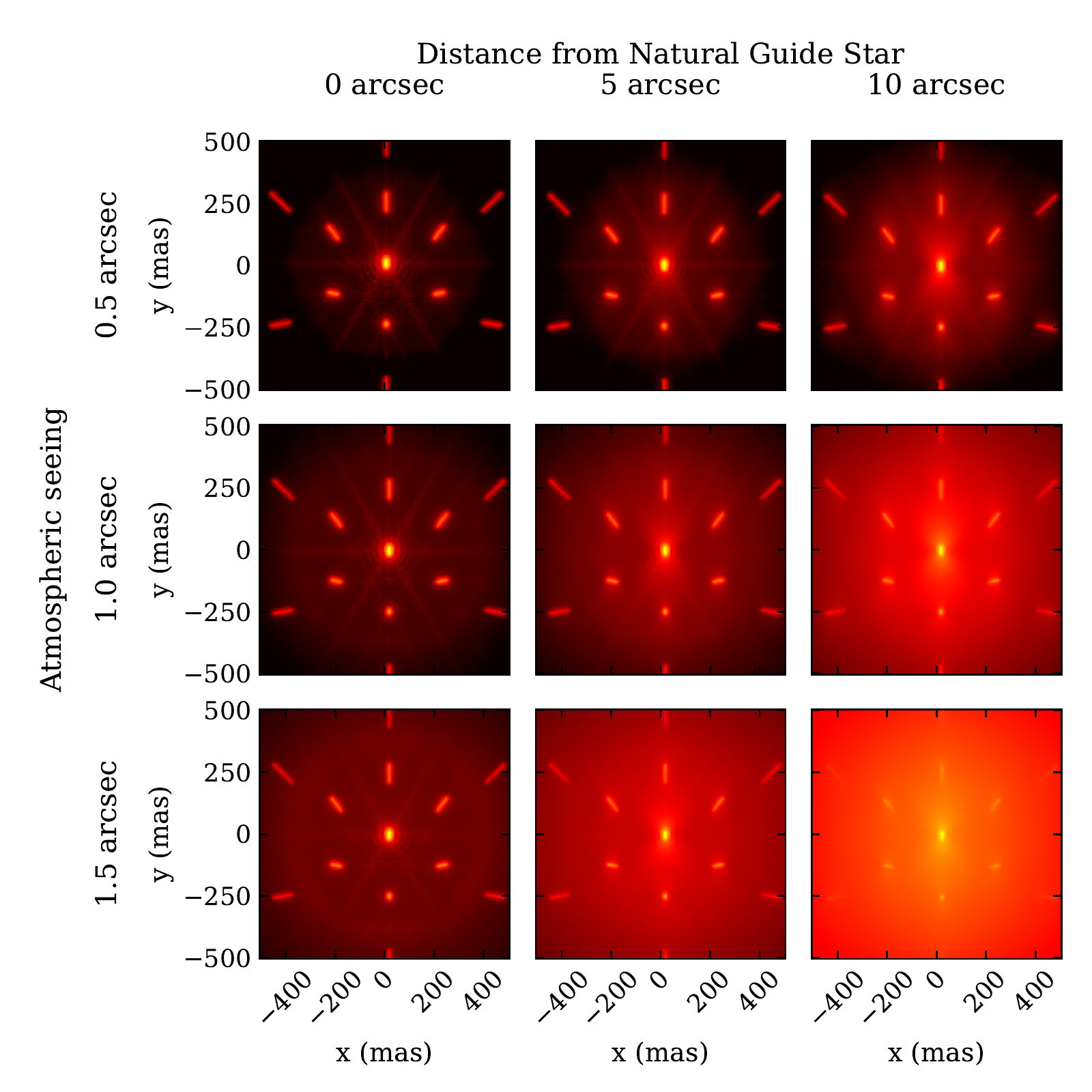}
    \caption{Comparison of simulated \textit{H}-band PSFs at three different seeing conditions and three different angular distances to the natural guide star used by the SCAO system. In all images $\Delta R = 30$ mas.}
    \label{fig:image_comparison}
\end{figure}

For the following we choose the \textit{Triangular apertures} mask design and use it in our simulation framework to assess the suitability of this mask for the performance validation of the MICADO ADC. We also explore the atmospheric conditions and observation prerequisites, so that we can recommend under what conditions to validate the ADC performance. The explored parameter space is summarised in Table~\ref{tab:par_space} and some of the corresponding PSFs are shown in Figure~\ref{fig:image_comparison}. 

\begin{table}[b]
    \caption{The parameter space that we explored in our simulations. Atmospheric turbulence and anisoplanatism have an impact on the contrast of the PSF core and the diffraction speckles relative to the background intensity. This impacts how well the dispersion in the image can be characterised.}
    \centering
    \begin{tabular}{lp{4cm}}
        \hline
        Parameter & Value \\
        \hline
        Passband & \textit{J} (1.15 - 1.345 \textmu m)\newline \textit{H} (1.49 - 1.78 \textmu m)\newline \textit{K\textsubscript{s}} (1.97 - 2.32 \textmu m) \\
        Atmospheric seeing (arcsec) & 0.1, 0.3, 0.5, 1.0, 1.5\\
        Distance from Natural Guide Star (arcsec) & 0, 5, 10\\
        Atmospheric dispersion (mas) & 0, 1, 2.5, 5, 10, 20, 30, 40,\newline 50, 60, 70, 80, 90, 100 \\
        \hline
    \end{tabular}
    \label{tab:par_space}
\end{table}

\begin{figure}
    \centering
    \includegraphics[width=0.45\linewidth]{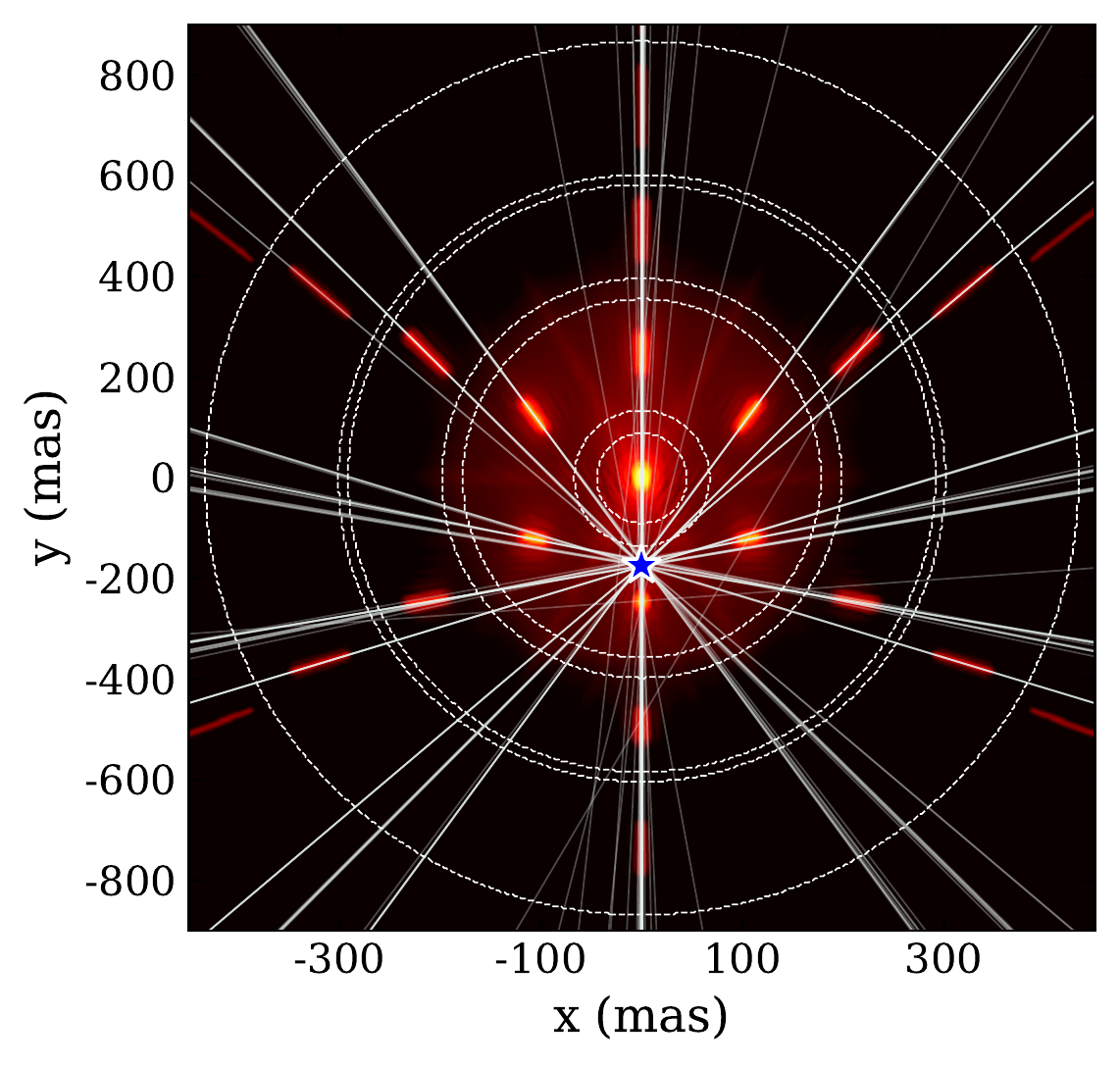}
    \caption{This \textit{H}-band image, with a NGS offset of 5 arcseconds and 0.5 arcseconds of seeing, illustrates the analysis steps that we take to determine the radiation centre. First, annuli are defined for each diffraction order, in which the artificial speckles are characterised using intensity based contours. These contours allow us to fit ellipses at fixed intensity levels, from which we can draw lines through the semi-major axes. Finally, the radiation centre is determined, here denoted by the blue star, and eq.~\eqref{eq:atm_disp_measure} is applied.}
    \label{fig:example_analysis}
\end{figure}

Analysis on the images largely follows Ref.~\citenum{vandenBorn2022} and is done through the following steps. First a set of annuli is defined, which contain in each annulus a single diffraction order. For each of these annuli, a different set of contour levels is chosen with sufficient signal to the local background intensity. Then, a contour plot is generated. The major axis of the best fitted ellipse through each sufficiently large contour defines a line. From these lines a least squares intersection point is determined. Now, the distance between the intersection point and the PSF core is known. Finally, we apply equation~\eqref{eq:atm_disp_measure} to find the chromatic elongation in the image. A visual representation of this process is given in Fig.~\ref{fig:example_analysis}.

\begin{figure}
    \centering
    \includegraphics[width=\linewidth]{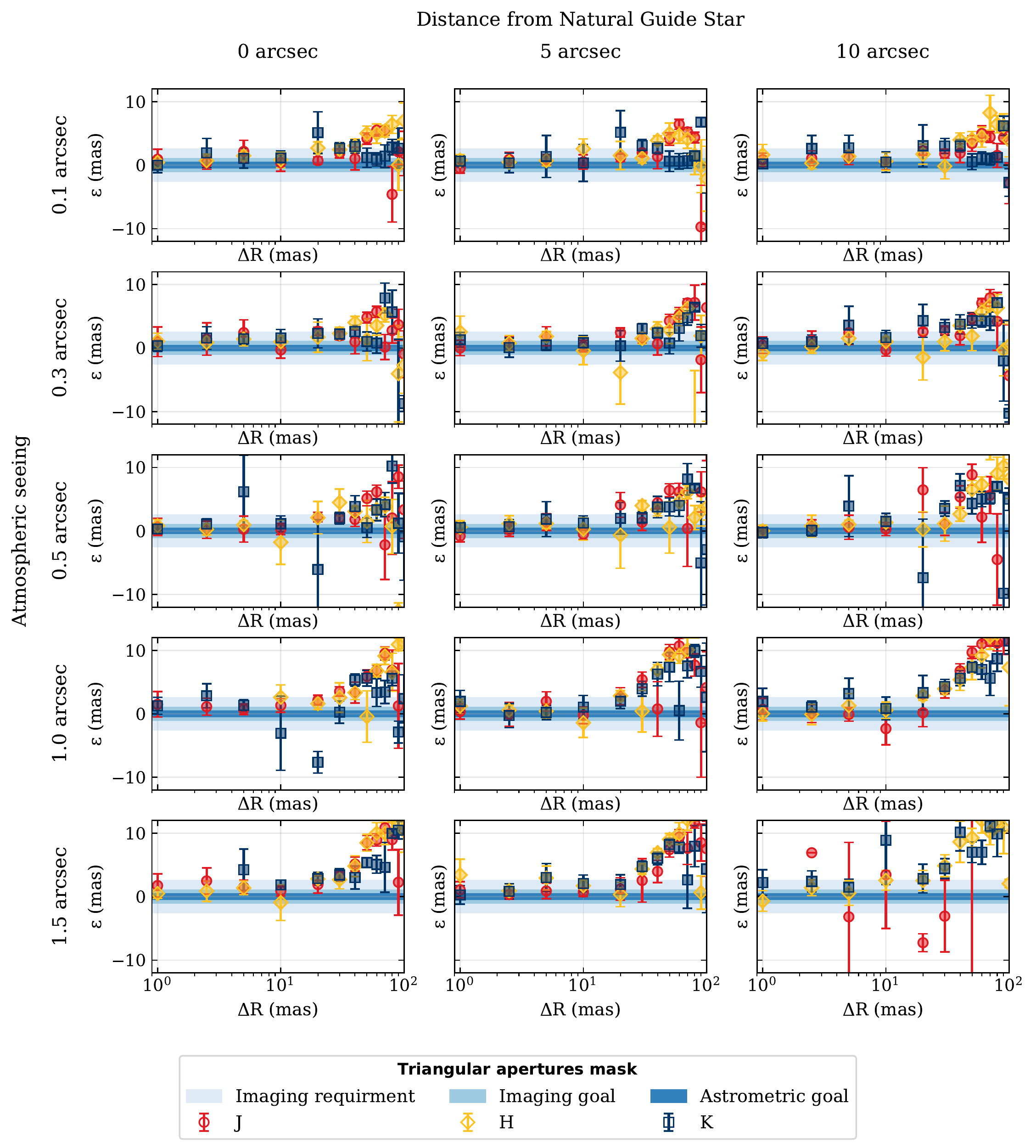}
    \caption{Shown here is the accuracy with which the true value of $\Delta R$ is recovered in our simulated images as a function of $\Delta R$. Generally, the dispersion in the images is recovered to within 2.5~mas of the true value, when the total dispersion is less than 10~mas. Worse observing conditions do have a limited impact on the accuracy. When it becomes difficult to distinguish the diffraction speckles, the measurement accuracy collapses completely, as seen in the bottom right panel.}
    \label{fig:speckle_method_results}
\end{figure}

This process is repeated for all images. Because we know exactly the true value of the dispersion in the image, we can inspect the accuracy with which it is recovered. This is shown in Fig.~\ref{fig:speckle_method_results}. We find that the error $\varepsilon$ between the true value and retrieved value of $\Delta R$ is generally within the requirement of 2.5~mas, when the true dispersion is less than 10 mas. As the dispersion magnitudes increase, the accuracy of the measurement decreases. Similarly, worse observing conditions make the measurement more difficult. In general, median seeing conditions at Cerro Armazones or better should be sufficient to validate the performance of the ADC. 

From Fig.~\ref{fig:speckle_method_results} it is yet unclear if the correction goal of 1.0~mas can be verified consistently. The astrometric goal of 0.4~mas will be difficult to conclusively validate with the present method. However, there are various improvements that can be implemented. First of all, the analysis here assumes that MICADO is using the standard imaging mode, which has a plate scale of 4~mas per pixel. The instrument also comes with a zoom mode, where a different set of optics is used after the pupil. In this mode the plate scale changes to from 4 to 1.5~mas per pixel. Also, the current analysis was very much a first attempt to assess the feasibility of the speckle method for MICADO. The use of manually defined contour masks is inefficient and the results are relatively sensitive to the provided input values. Better analysis methods are sure to exist. For example, in Ref.~\citenum{Pathak2016} a very similar approach is used to position an ADC using a closed-loop control system, where they find the radiation centre in an iterative manner. Alternatively, the visual extent of the dispersion in the image could prove well suited for a machine learning algorithm.

\subsection{Alternative calibration methods}
Up to now, there has been no standard method for the validation and calibration of ADCs. Here, we will mention a few alternatives to the speckle method and shortly discuss their advantages and drawbacks. 

Perhaps the closest thing to an industry standard method is through the direct analysis of the PSF shape. By altering the ADC configuration the elongation of the PSF changes and an optimum position can be determined. Interpolation between the data points can allow for an accurate determination of the point of minimum residual dispersion. See for example Ref.~\citenum{Cabral2021}. When the PSF elongation due to chromatic dispersion is small, the likelihood increases that the elongation becomes dominated by other systematic effects, such as geometric distortions, telescope vibrations or optical aberrations. To counteract this issue, one could find the configuration of the ADC for which the residual dispersion is at a maximum. Assuming a rotating ADC, the best prism orientation is then obtained by rotating one of the prism pairs by exactly 180\degr.

The above method requires a substantial number of on-sky observations to explore the parameter space in which the ADC must perform. With MICADO and the ELT this may be an expensive endeavour. However, it can always be used as a fallback option.

A different approach, requiring no on-sky observations at all, is to rely on look-up tables created from the instrument optical model and a model for the atmosphere\cite{Kopon2013}. Unfortunately, with the increase in resolution offered by the ELT, the accuracy of the assumed atmosphere models may be questioned\cite{Spano2014, vandenBorn2020a} and thus this method may not be sufficient in the early phases of ELT operations. A partial solution to this issue would be the characterisation of the atmospheric dispersion through a spectroscopic measurement. With the single slit spectroscopic functionality offered by MICADO, the shift in the spectroscopic trace on the detector could be used to validate or update the used atmospheric dispersion model\cite{Skemer2009,Wehbe2020,Wehbe2021}. The ADC positioning model would then be updated accordingly.

Finally, a new method is described in Ref.~\citenum{Pierik2022}, where the ADC position is optimised through the minimisation of the colour-dependent offsets with respect to a reference astrometric solution. This method appears to work well, even for observing sites where the atmospheric conditions are sub-optimal. But again, the increase in resolution and sensitivity provided by the 39 meter primary aperture of the ELT will make the number of reference astrometric fields, with a sufficient number of astrometrically accurate sources within the field of view, fairly limited. This will limit the extent to which the MICADO ADC can then be characterised using this method.

The speckle method that we have discussed in detail in this work has some advantages and disadvantages as well. Inherent to the amplitude modulation nature of the mask design, about half of the light will be blocked by the mask. For the validation of the ADC performance in MICADO this should not pose a problem, as a sufficiently bright point source can be selected. If anything, this will decrease the likelihood of source confusion as the signal of fainter objects will be much lower. The location of the mask within the cold instrument allows for daytime calibration if a broadband calibration light source is used. Earlier analyses on the optical design of MICADO have suggested that internal transmissive optics, as well as the ADC optics themselves, introduce a small amount of field varying chromatic elongation. This elongation would be on the order of 1~mas and cannot be corrected for over the full imaged field. Because the ADC calibration mask is located within the optical pupil, the diffraction speckles affect each source in the image. As a result, it is possible to measure the residual dispersion as a function of field position. If the accuracy of the dispersion retrieval, discussed in section~\ref{subsec:validation}, can be improved upon, then it should be possible to detect this effect.

Perhaps the main drawback of the speckle method is that it requires a dedicated pupil optic. In principle, one of the deformable mirrors can be used as well, but this limits the design of the PSF shape through the limited spatial frequencies and patterns that the deformable mirror can project. 

%% file: 06_conclusion.tex
\section{CONCLUSION}\label{sec:conclusion}
Although the concept of an atmospheric dispersion is not new, the advent of upcoming extremely large telescopes requires some reexamination on the correction of atmospheric dispersion. First of all, the promised increase in resolution makes an ADC an absolute necessity well into the near-infrared for any imager aiming to be diffraction limited. This also means that it is no longer sufficient to place the prisms of the ADC in an approximately correct position. Furthermore, the design of the MICADO ADC is made more difficult due to the cryogenic environment it will operate in. 

In this report, we have discussed the optical and mechanical design of the MICADO ADC, including how the cold environment affects the performance. The refractive index of the S-FPL51 and S-LAH71 glasses at 80~K differs from the refractivity at room temperature as provided by OHARA. An additional design optimisation will be necessary to maximise the ADC performance. Likewise, to update the mechanical design the coefficient of thermal expansion of these glasses was measured. We have discussed the prism mounting solution that can accommodate for these shrinkages.

The friction drive concept that will position the Amici prisms has been discussed. Through several tests, we have shown that such a concept can work for a cryogenic ADC and that a sufficient endurance of the mechanism can be obtained. 

Finally, we have laid out a promising performance validation method. The increase in sensitivity and resolution of the ELT, combined with the challenging requirements on the instrument, makes the validation and calibration of the MICADO ADC difficult. Methods that might work on smaller telescope will not necessarily be sufficiently accurate or time efficient. Therefore, we investigated the feasibility of a method that employs a diffraction mask, which visually magnifies any chromatic elongation present in the PSF. A diffraction mask, consisting of a repeating set of triangular apertures provides the best trade-off between design simplicity and speckle contrast. A wide range of PSF simulations was performed to determine that a residual dispersion of around 2.5~mas can still be distinguished. Accordingly, this method will allow for a quick quantitative assessment of the ADC performance with only a single image.